# Chain-Based Representations
# for Solid and Physical Modeling

Antonio DiCarlo, Franco Milicchio, Alberto Paoluzzi, and Vadim Shapiro

*Abstract*— In this paper we show that the (co)chain complex associated with a decomposition of the computational domain, commonly called a mesh in computational science and engineering, can be represented by a block-bidiagonal matrix that we call the Hasse matrix. Moreover, we show that topology-preserving mesh refinements, produced by the action of (the simplest) Euler operators, can be reduced to multilinear transformations of the Hasse matrix representing the complex. Our main result is a new representation of the (co)chain complex underlying field computations, a representation that provides new insights into the transformations induced by local mesh refinements. Our approach is based on first principles and is general in that it applies to most representational domains that can be characterized as cell complexes, without any restrictions on their type, dimension, codimension, orientability, manifoldness, connectedness.

## I. Introduction

### A. Motivation

The boundary representation has become the representation of choice in many academic and virtually all commercial solid modeling systems. As a consequence, most geometric, scientific and engineering applications have to be formulated in terms of boundary representations, often leading to nontrivial representation conversion problems. Well known examples of such problems include Boolean set operations, finite element meshing, and subdivision algorithms.

Formally, all boundary representations are widely recognized as graph-based data structures [1]–[4] representing one of several possible cells complexes [5]–[8]. Space requirements and computational efficiency of such data structures have been studied in the literature (see, *e.g.*, [9]). Historically, such cell complexes have been restricted to (unions of) two-dimensional orientable manifolds, but a number of extensions to more general orientable cellular spaces have been proposed (see, *e.g.*, [8], [10], [11]). Depending on a particular choice of data structures, boundary representations are constructed, edited, and updated using a small set of basic operators on the graph representation, while preserving and/or updating the basic topological invariants of the cell complex. Such operators are commonly called Euler operators [3], [10], [12], because they enforce the Euler-Poincaré formula. All higher-level algorithms and applications of boundary representations are implemented in terms of such operators.

This evolutionary development of boundary representations also led to several fundamental difficulties:

- Variety of assumptions about the cell complexes and graph representations make standardization difficult. This in turns complicates the issues of data exchange and transfer, and leads to proliferation of incompatible algorithms.
- Boundary representation algorithms are dominated by graph searching algorithms (boundary traversals) that tend to force serial processing. Nor is it clear how to combine such graph representations with multi-resolution representations and algorithms.
- Extending boundary representations to more general cellular spaces has proved challenging. Despite many proposals, most commercial systems are still restricted to two-dimensional orientable surfaces.
- Last, but not least, solid modeling has developed into a highly specialized discipline that is largely disconnected from many standard computational techniques. In particular, boundary representations do not appear to be directly related to the methods for physical analysis and simulation such as finite differences, finite elements, and finite volumes.

In this paper, we claim that *all* representations of cell complexes are properly represented by a (co)chain complex [13], [14]. It captures all combinatorial relationship of interest in solid and physical modeling formally and unambiguously, using standard algebraic topological operators of boundary $\partial$ and coboundary $\delta$. We show that the (co)chain complex can be represented by a sparse block-bidiagonal matrix that we call the Hasse matrix. We also show that topology-preserving refinements of such cell complexes correspond to simple Euler operators and are easily formulated as multi-linear transformations of the Hasse matrix. There are at least three important consequences of our proposal. First, the proposed approach applies to all cell complexes, without restrictions on type, dimension, codimension, orientability, manifoldness, and so on. Secondly, as we will see in Section VIII, this formalism unifies geometric and physical computations within a common formal computational structure. And finally, our formulation explicitly shows that many geometric computations (and computations with boundary representations in particular) can be formulated and implemented in terms of standard sparse matrix computational techniques, opening a possibility for a wide range of computational breakthroughs and opportunities.

### B. Related Work

Algebraic-topological properties of boundary representations are well understood—see [3], [5], [8], [15] for details. Branin [16] and Tonti [17] advocated a unified discrete view of all physical theories using concepts from algebraic topology and the de Rham cohomology. More recently, this early research led to new efforts in developing unified computational models and languages for analysis, simulation, and engineering design. Notably, Palmer and Shapiro [18] proposed a unified computational model of engineering systems that relies on concepts from algebraic topology. A number of researchers went beyond the use of chains and cochains as general-purpose data types, considering that a sound numerical method should reflect the algebraic-topological structure of the underlying physical theory in a faithful way. Notably, Strang [19] observed that the FEM encodes a pervasive balance pattern, which is at the center of the classification in [17]. Mattiussi [20] provided interpretations of FEM, FVM, and FDM in terms of the topological properties of the corresponding field theory. Tonti [21] presented his cell method as a direct discrete method, bypassing the underlying continuum model. In [22] FDMs that satisfy desired topological properties are discussed. In our previous work [23], physical information is attached to an adaptive, full-dimensional decomposition of the computational domain. Giving pre-eminence to the cells of highest dimension allows to generate the geometry and to simulate the physics simultaneously. Such a formulation removes artificial constraints on the shape of discrete elements and unifies commonly unrelated finite methods into a single computational framework [24]. Our goal is to graft this approach to field modeling on an already established computational framework for geometric modeling with cell complexes [25]. This framework has been recently extended to provide parallel and progressive generation of very large datasets using streams of continuous approximations of the domain with convex cells [26]. The approach also supports progressive Boolean operations [27], providing continuous streaming of geometrical features and adaptive refinement of their details.

### C. Overview

This paper is a revised and augmented version of our contribution to the 2007 ACM Symposium on Solid and Physical Modeling [28], that now explicitly accounts for unit (co)chains, cell size and inner product. We also made an effort to render the present version more palatable for the prospective reader, and as self-contained as possible. All technical notions—apart from the most basic ones, listed at the beginning of Section II—are explicitly defined and explained through examples.

In Section II we review the basic notions of (co)chain and (co)boundary and their representations using matrices and the Hasse diagram. We make clear that chains with real coefficients serve to render cells measurable; correspondingly, cochains represent densities of physical quantities with respect to cell measures. The duality between boundary and coboundary operators and its representation through matrix transposition are highlighted. Section III introduces our block-matrix representation of a (co)chain complex. In Section IV we use algebraic-topological notions to define a minimal set of operators as transformations between cell complexes that preserve the Euler characteristics. These operators are shown to correspond to multilinear transformations of the Hasse matrix in Section V. Section VI demonstrates how common algorithms for splitting a cell complex may be formulated in algebraic-topological terms. Section VII explains how the proposed representation may unify geometric and physical modeling in a common computational framework, showing how local adjacency information—leading to the discrete Laplace-deRham operators—is straightforwardly obtained through simple linear algebra, after chains have been identified with (real-valued) cochains. The central role played here (and only here) by a further independent mathematical structure—the inner product between (co)chains—is brought to the fore in Sections VII and VIII. Section IX provides some concluding remarks.

## II. Background

We take for granted the elementary notions of *simplex*, *cell*, *orientation*, *cell complex*, *face*, and refer the reader to [5], [13]. (In Fig. 1, $\sigma_2^2, \sigma_2^3$, and $\sigma_2^4$ are simplices, $\sigma_2^1$ is not; the faces of $\sigma_2^1$ are $\sigma_1^1, \sigma_1^2, \sigma_1^6$, and $\sigma_1^3$.) Apart from 0-cells, we take all cells as *relatively open*.

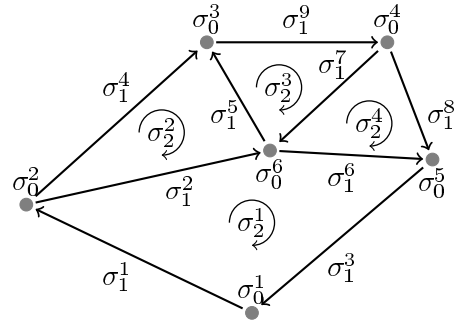

Fig. 1. A 2-complex $K$. Notice that its 2-cells are coherently oriented.

### A. Chains and Cochains

Let $K$ be a cell complex representing a *finite* partition of a compact set $D \subset \mathbb{E}^d$, the $d$-dimensional Euclidean point space (or, more generally, $D \subset \mathcal{M}$, a Riemannian manifold modeled on $\mathbb{E}^d$). By way of illustration, Fig. 1 exhibits a small 2-complex partitioning a polygonal subset of $\mathbb{E}^2$. We call *p-skeleton* $K_p \subset K$ the subset of oriented $p$-cells of $K$, and denote with $k_p$ the number of $p$-cells: $k_p := \#K_p$, hence

$$\#K = k_0 + k_1 + \cdots + k_d.$$

For the 2-complex $K$ in Fig. 1, *e.g.*, $\#K = 6+9+4$. The $p$-skeleton $K_p$ will be ordered by labeling each $p$-cell $\sigma_p$ with a positive integer: $K_p = (\sigma_p^1, \ldots, \sigma_p^{k_p})$. In the following, the ordinal and/or the dimension of cells will be dropped from notation whenever convenient, and the complex $K$ will be identified with the tuple of its ordered $p$-skeletons: $K \cong (K_0, \ldots, K_d)$.

*1) Chains:* Let $(G, +)$ be a nontrivial *abelian* (*i.e.*, commutative) *group*. A *$p$-chain* of $K$ with coefficients in $G$ is a mapping $c_p : K_p \to G$ such that, for each $\sigma \in K_p$, reversing a cell orientation changes the sign of the chain value:

$$c_p(-\sigma) = -c_p(\sigma) \,.$$

*Chain addition* is defined by addition of chain values: if $c_p^1, c_p^2$ are $p$-chains, then $(c_p^1 + c_p^2)(\sigma) = c_p^1(\sigma) + c_p^2(\sigma)$, for each $\sigma \in K_p$. The resulting group is denoted $C_p(K; G)$. When clear from the context, the group $G$ is often left implied, writing $C_p(K)$.

Let $\sigma$ be an oriented cell in $K$ and $g \in G$. The *elementary chain* whose value is $g$ on $\sigma$, $-g$ on $-\sigma$ and $0$ on any other cell in $K$ is denoted $g\sigma$. Each chain can then be written in a unique way as a (finite) sum of elementary chains:

$$c_p = \sum_{k=1}^{k_p} g_k \, \sigma_p^k \,. \tag{1}$$

Chains are often thought of as attaching orientation and multiplicity to cells: if coefficients are taken from the smallest nontrivial group, *i.e.*, $G = \{-1, 0, 1\}$, then cells can only be *discarded* or *selected*, possibly *inverting their orientation*. However, we are mainly interested in *real*-valued chains, and strongly rely on the extra structure imparted to the set $\mathbb{R}$ of real numbers by *multiplication*. In fact, the ability to add (and subtract) chains is not enough: the very idea of mesh refinement (and coarsening) requires chains to be also *scaled*. Hence the full *linear* structure is brought to bear, as argued below.

Real-valued chains attach a signed $p$-measure to $p$-cells—such as length to 1-cells, area to 2-cells, volume to 3-cells, thus restoring part of the geometrical information left out by the purely topological construction of a cell complex. Thanks to the multiplicative structure of the real field, all $\mathbb{R}$-valued elementary chains on the $p$-cell $\sigma_p^k$ are multiples (or submultiples) of the *unit chain* $u_p^k$, whose value is $1$ on $\sigma_p^k$, $-1$ on $-\sigma_p^k$ and $0$ on any other cell in $K_p$. Therefore, each chain $c_p$ can be seen as a *linear combination* of unit chains, and $C_p(K; \mathbb{R})$ as a *linear space* over $\mathbb{R}$, spanned by the set of unit $p$-chains, which constitutes its *standard basis*:

$$c_p = \sum_{k=1}^{k_p} \lambda_k \, u_p^k \,. \tag{2}$$

Notice that representation (1) would formally coincide with (2), if each cell $\sigma_p^k$ were identified with the corresponding unit chain $u_p^k$. However, this seemingly innocuous move should definitely be avoided, being tantamount to assigning one and the same size to all cells. On the contrary, we want to tune size freely, by identifying each cell with a selected elementary chain, *i.e.*, a multiple of the corresponding unit chain:

$$\sigma_p^k \cong \mu_p^k \, u_p^k \,, \tag{3}$$

the scalar $\mu_p^k$ being the *size* imparted to the cell $\sigma_p^k$. The size $\mu_p^k$ is assumed to be *positive* (orientation being accounted for by $u_p^k$) and significantly different from zero:[1] $\mu_p^k \gg 0$. 0-cells are systematically given a unit size: $\mu_0^k = 1 \;\Leftrightarrow\; \sigma_0^k \cong u_0^k$. Our motivation is illustrated in the following straightforward example.

[1] In Section II-B.5 cell sizes will be used as divisors. Cf. also the definition of $\operatorname{sgn}_\varepsilon$ in Section VI-A.

*Example 1 (Measuring cell size):* Three different partitions of the same 1-D interval are arranged in Fig. 2. The upper and middle ones are indistinguishable as *cell* complexes, *i.e.*, on mere topological grounds. As *chain* complexes, however, they may differ: their two elementary 1-chains, if meant to represent *(cell length)/(interval length)*, take the values $(.5, .5)$ for the upper mesh and $(.75, .25)$ for the middle one. The lower mesh is a *refinement* of the middle one: its four 1-chains take the values $(.5, .25, .125, .125)$.

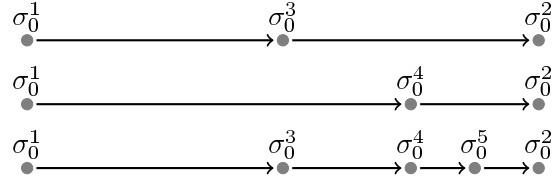

Fig. 2. Three 1D meshes.

*2) Cochains:* By definition, the group of *$p$-cochains* of $K$ with coefficients in $G$ is the group of *homomorphisms* of $C_p(K; G)$ into $G$:

$$C^p(K; G) \coloneqq Hom(C_p(K; G), G),$$

*i.e.*, $\gamma^p \in C^p(K; G)$ if and only if $\gamma^p : C_p(K; G) \to G$ and

$$\gamma^p \Big( \sum_{k=1}^{k_p} g_k \, \sigma_p^k \Big) = \sum_{k=1}^{k_p} \gamma^p \Big( g_k \, \sigma_p^k \Big)$$

for all $(g_1, \dots, g_{k_p}) \in G^{k_p}$.

If (co)chains are real-valued, then $\gamma^p(\lambda \, u_p) = \lambda \, \gamma^p(u_p)$, and each cochain $\gamma^p$ can be seen as a linear combination of the *unit $p$-cochains* $\eta_1^p, \dots, \eta_{k_p}^p$, whose value is $1$ on a unit $p$-chain and $0$ on all others:

$$\gamma^p = \sum_{k=1}^{k_p} \lambda^k \, \eta_k^p \,, \quad \text{where} \quad \eta_j^p(u_p^i) = \begin{cases} 1 & \text{if } i = j \,; \\ 0 & \text{otherwise.} \end{cases} \tag{4}$$

Therefore, $C^p(K; \mathbb{R})$ can be seen as the *dual space* of $C_p(K; \mathbb{R})$. It is spanned by the set of unit $p$-cochains, which constitutes the basis *dual* to the standard basis of $C_p(K; \mathbb{R})$. The evaluation of a real-valued cochain is aptly denoted as a duality pairing, in order to stress its bilinear property:

$$\gamma^p(c_p) = \langle \gamma^p, c_p \rangle.$$

Real-valued cochains represent *densities* with respect to the measures imparted to cells by real-valued chains, and the above duality pairing is a discrete preliminary to integration: its value yields the integral of the density $\gamma^p$ over the chain $c_p$, *i.e.*, the content in the support of $c_p$ of the quantity gauged by $\gamma^p$.

### B. Boundary and Coboundary

*1) Boundary Operator:* The *boundary operator*

$$\partial_p : C_p(K) \to C_{p-1}(K)$$

is first defined on *simplices* (see [13, Ch. 1 §5]). If $\sigma_p$ is the oriented $p$-simplex spanned by the ordered set of $p+1$ points $(\boldsymbol{x}_0, \ldots, \boldsymbol{x}_p)$, then

$$\partial_p \,\sigma_p := \sum_{k=0}^{p} (-1)^k \sigma_{p-1,\,k}\,,$$

where $\sigma_{p-1,\,k}$ is the $k$-th *face* of $\sigma_p$, *i.e.*, the oriented $(p-1)$-simplex spanned by $(\boldsymbol{x}_0, \ldots, \boldsymbol{x}_{k-1}, \boldsymbol{x}_{k+1}, \ldots, \boldsymbol{x}_p)$. *E.g.*, with reference to Fig. 1, we have $\partial_2 \,\sigma_2^2 = -\sigma_1^2 - \sigma_1^5 + \sigma_1^4$.

The next step is to extend $\partial_p$ to *cells* by partitioning them into simplices, and exploiting the additivity of the $\partial_p$ operator. *E.g.*, the nonsimplicial 2-cell $\sigma_2^1$ in Fig. 1 may be partitioned into (at least) two 2-simplices. Of course, partitioning is not unique, but the result does *not* depend on the partition used, since the contributions of all faces internal to the partitioned cell cancel out (see [13, Ch. 4 §39]). For the nonsimplicial cell in Fig. 1, one gets $\partial_2 \,\sigma_2^1 = \sigma_1^1 + \sigma_1^2 + \sigma_1^6 + \sigma_1^3$.

The boundary operator is then extended to elementary chains, by taking

$$\partial_p(g\sigma) := g(\partial_p \sigma)\,,$$

and finally to all chains by additivity. In conclusion, the boundary operator is a *homomorphism*:

$$\partial_p \in Hom(C_p(K), C_{p-1}(K)).$$

In particular, it preserves the *linear* structure real-valued chains have been endowed with. In the following (see Section II-B.5), we shall represent $\partial_p$ by means of the $k_{p-1} \times k_p$ matrix $[\partial_p]$ of its components with respect to the standard bases of $C_p(K)$ and $C_{p-1}(K)$.

*2) Coboundary Operator:* The *coboundary operator*

$$\delta^p : C^p(K) \to C^{p+1}(K)$$

acts on $p$-cochains as the *dual* of the boundary operator $\partial_{p+1}$ on $(p+1)$-chains: for all $\gamma^p \in C^p$ and $c_{p+1} \in C_{p+1}$,

$$\langle \delta^p \gamma^p, c_{p+1} \rangle_{p+1} = \langle \gamma^p, \partial_{p+1}\, c_{p+1} \rangle_p\,.$$

For once, we have labelled each duality bracket with the dimension of the (co)chains it acts on. Recalling that chain-cochain duality means integration, the reader will recognize this defining property as the combinatorial archetype of *Stokes' theorem*. Denoting the dual of an operator by starring its symbol, we shall write:

$$\delta^p = \partial_{p+1}^*\,.$$

By definition, the coboundary operator is a homomorphism preserving the linear structure real-valued cochains have been endowed with:

$$\delta^p \in Hom(C^p(K), C^{p+1}(K)).$$

In the following (see Section II-B.5), we shall represent $\delta^p$ by means of the $k_{p+1} \times k_p$ matrix $[\delta^p]$ of its components with respect to the standard bases of $C^p(K)$ and $C^{p+1}(K)$, *i.e.*, the bases dual to the standard bases of $C_p(K)$ and $C_{p+1}(K)$. Since we use *dual* bases, matrices representing *dual* operators are the *transpose* of each other: for all $p = 0, \ldots, d-1$,

$$[\delta^p] = [\partial_{p+1}]^t.$$

*Example 2 (Cochains, Coboundary and Electrostatics):* Consider once again the 2-complex $K$ in Fig. 1, its support $[\![K]\!]$ (*i.e.*, the point set union of all cells in $K$), and a scalar field $\phi : [\![\mathbb{K}]\!] \to \mathbb{R}$, to be thought of as the *electrostatic potential*. The discrete counterpart of $\phi$ is the 0-cochain $\phi^0$ sampling it at the six 0-cells of $K$. The *electric field* $\mathbf{E}$ is the negative gradient of the electrostatic potential: $\mathbf{E} = -\nabla \phi$. A measurable quantity is its *circulation* along any curve $\mathfrak{c} : [0,1] \to [\![\mathbb{K}]\!]$, defined as

$$C(\mathbf{E}, \mathfrak{c}) := \int_0^1 \mathbf{E}(\mathfrak{c}(s)) \cdot \mathfrak{c}'(s)\, ds\,,$$

where a prime denotes differentiation with respect to the curvilinear coordinate. Since $\mathbf{E} = -\nabla\phi$, it turns out that

$$C(\mathbf{E}, \mathfrak{c}) = -\int_0^1 \nabla\phi|_{\mathfrak{c}(s)} \cdot \mathfrak{c}'(s)\, ds = \phi(\mathfrak{c}(0)) - \phi(\mathfrak{c}(1))\,.$$

The analog in $K$ of a curve is a 1-chain, and the discrete counterpart of the vector field $\mathbf{E}$ is the 1-cochain $C^1$ sampling its circulation along the nine 1-cells of $K$, delivered by the opposite of the coboundary of $\phi^0$:

$$C^1 = -\delta^0 \phi^0,$$

*i.e.*, $\delta^0$ is the discrete counterpart of $\nabla$ acting on scalar fields. Since $\mathbf{E}$ admits a potential, its circulation along the boundary of any disk $\mathcal{D} \subset [\![\mathbb{K}]\!]$ is null:

$$C(\mathbf{E}, \partial\mathcal{D}) = \int_{\mathcal{D}} \mathrm{rot}\, \mathbf{E} = 0\,.$$

Analogously, for any linear combination of the four 2-cells in $K$, *i.e.*, for any 2-chain $c_2 \in K_2$, we have

$$\langle C^1, \partial_2\, c_2 \rangle = \langle \delta^1 C^1, c_2 \rangle = 0\,,$$

since $\delta^1 \circ \delta^0 = 0$ (cf. Section II-C.1). Clearly, $\delta^1$ is the discrete counterpart of the curl operator on vector fields.

*3) Matrix Representation of Chains and Cochains:* Since $p$-chains and $p$-cochains form dual linear spaces (cf. Section II-A), they lend themselves to the usual representation of vectors and covectors as column and row matrices. The standard basis of $C_p$ (cf. Section II-A.1) and the dual basis of $C^p$ (cf. Section II-A.2) will be used throughout.

The components $g_1, \ldots, g_{k_p}$ of a $p$-chain $c_p$ in (2) may be organized into a *column* matrix $\mathbf{c}_p = [c_p]$. Analogously, the components $g^1, \ldots, g^{k_p}$ of a $p$-cochain $\gamma^p$ in (4) may be organized into a *row* matrix $\mathbf{y}^p = [\gamma^p]$. The duality pairing between $\gamma^p$ and $c_p$ is represented by a matrix product:

$$\langle \gamma^p, c_p \rangle = [\gamma^p][c_p].$$

*4) Incidence Matrices:* The intersection between $p$-cells and *relatively closed* $(p+1)$-cells may be characterized by the $p$-incidence matrix $\mathbf{B}_p$, whose entries $B_p^{ij}$ are defined as follows:

$B_p^{ij} = 0$ if $\sigma_p^i \cap \overline{\sigma}_{p+1}^j = \emptyset$, $\overline{\sigma}$ being the *closure* of $\sigma$;

$B_p^{ij} = \pm 1$ otherwise, with $+1$ ($-1$) if the orientation of $\sigma_p^i$ is equal (opposite) to that of the corresponding face of $\sigma_{p+1}^j$.

Be it noted that the information stored in $\mathbf{B}_p$ is purely topological. As is obvious, its transpose $\mathbf{B}_p^t$ describes how $(p+1)$-cells intersect with $p$-cells.

*5) Matrix Representation of Boundary and Coboundary Operators:* We now introduce the notion of *measured p-incidence*, by defining a matrix $\mathbf{M}_p$ whose entries $M_p^{ij}$ depend on both topology (through $p$-incidence) and measure (through cell size):

$$M_p^{ij} := (\mu_p^i / \mu_{p+1}^j) B_p^{ij}.$$

It should be no surprise that $\mathbf{M}_p$ is exactly the matrix representing the boundary operator $\partial_{p+1}$ with respect to the standard bases of $C_{p+1}$ and $C_p$:

$$[\partial_{p+1}] = \mathbf{M}_p. \tag{5}$$

Consequently (cf. Section II-B.2), its transpose represents the coboundary operator $\delta^p$ with respect to the dual bases:

$$[\delta^p] = \mathbf{M}_p^t. \tag{6}$$

If size represents length, area, volume for 1-, 2-, and 3-cells, respectively, then measured-incidence matrices have physical dimension ($length$)$^{-1}$. Hence, $[\partial]$ and $[\delta]$ act as *first-order difference* operators.

Measure may be readily filtered out of $\mathbf{M}_p$, and $\mathbf{B}_p$ recovered, whenever needed:

$$B_p^{ij} = \mathrm{sgn}(M_p^{ij}).$$

### C. Hasse Diagram of a Chain Complex

Hasse diagrams, named after the German mathematician Helmut Hasse (1898–1979), illustrate the cover relation of a partial order, and are commonly used for representing partially ordered sets—or *posets*, for short.

If $N$ is a poset, its *Hasse diagram* is the graph $\mathcal{H} = (N, E)$, such that for any $x, y \in N$, there exists $(x, y) \in E$ if and only if $x < y$, and there is no $z \in N$ such that $x < z < y$. Given a $d$-complex $K$, let the sets $N, E$ be defined as follows:

1) $N := K_0 \cup K_1 \cup \cdots \cup K_d$,
2) $E := E_1 \cup \cdots \cup E_d$, with
3) $E_p := \{(\sigma_p, \sigma_{p-1}) \mid \sigma_{p-1} \in \partial \sigma_p\}$, $1 \leq p \leq d$,

then the graph $\mathcal{H}(K) := (N, E)$ provides a complete representation of $K$. Attaching a label from $\{+1, -1\}$ to the arc $(x, y) \in E_p$, denoted $\mathrm{sgn}(x, y)$, suffices to specify the relative orientation between the $p$-cell represented by the node $x$ and the $(p-1)$-cell represented by the node $y$. A (strictly positive) real label is needed to specify the size ratio $\mu_{p-1}(x) / \mu_p(y)$.

Given $\mathcal{H}(K) = (N, E)$, for each $x \in N$ define the subset of its "children" $N^x \subset N$:

1) $E^x := \{(x, y) \mid y \in N, (x, y) \in E\}$,
2) $N^x := \{y \mid y \in N, (x, y) \in E^x\}$.

Then, the boundary of the unit chain $u(x)$ represented by the node $x$ may be computed as:

$$\partial u(x) = \sum_{y \in N^x} \mathrm{sgn}(x, y) \, u(y),$$

where $u(y)$ denotes the unit chain represented by the node $y$. The boundary of a general chain follows by linearity.

*Example 3 (Hasse Diagram):* Fig. 3 represents a 2-complex embedded in $\mathbb{E}^2$, comprising two 2-cells, seven 1-cells and six 0-cells. Fig. 4 depicts its Hasse graph. The action of the boundary operators $\partial_1$ and $\partial_2$ and of their duals, the coboundary operators $\delta^0 = \partial_1^*$ and $\delta^1 = \partial_2^*$ is illustrated by arrows.

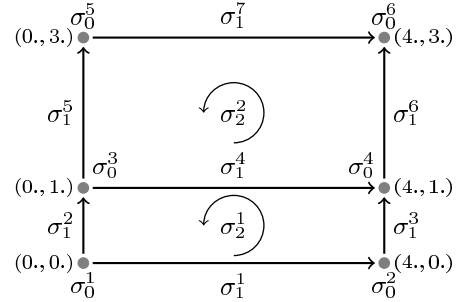

Fig. 3. A small 2-complex. Ordered pairs are Cartesian coordinates in $\mathbb{E}^2$.

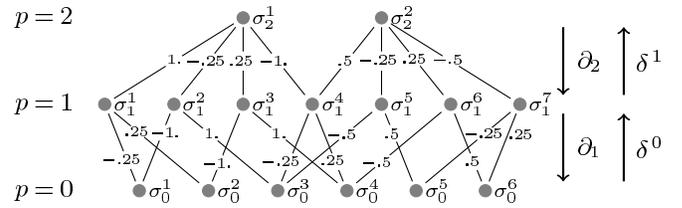

Fig. 4. The Hasse diagram of the 2-complex in Fig. 3. Each arc $(x, y)$ bears a non-null real label $\lambda(x, y)$, whose sign specifies the relative orientation between $x$ and $y$, while its modulus $|\lambda(x, y)|$ specifies the size ratio $\mu_{p-1}(y)/\mu_p(x)$.

*1) Chain and Cochain Complexes:* An abstract *chain complex* $\mathcal{C} = (C_p, \partial_p)$ is defined as a sequence of abelian groups $C_p$, paired with homomorphisms $\partial_p$, $p \geq 1$, satisfying the relations $\partial_p \circ \partial_{p+1} = 0$, for each $p \geq 1$:

$$\cdots \longrightarrow C_{p+1} \xrightarrow{\partial_{p+1}} C_p \xrightarrow{\partial_p} C_{p-1} \longrightarrow \cdots \longrightarrow C_1 \xrightarrow{\partial_1} C_0.$$

Analogously, an abstract *cochain complex* $\mathcal{C}' = (C^p, \delta^p)$ is a sequence of abelian groups $C^p$, paired with homomorphisms $\delta^p$, $p \geq 1$, satisfying the relations $\delta^p \circ \delta^{p-1} = 0$ ($p \geq 1$):

$$\cdots \longleftarrow C^{p+1} \xleftarrow{\delta^p} C^p \xleftarrow{\delta^{p-1}} C^{p-1} \longleftarrow \cdots \longleftarrow C^1 \xleftarrow{\delta^0} C^0.$$

If a chain complex $\mathcal{C}$ consists of *linear* spaces $C_p$ and *linear* operators $\partial_p$, then the dual spaces $C^p = C_p^*$ and the dual operators $\delta^p = \partial_{p+1}^*$ make up a cochain complex ($\mathcal{C}^*$).

*2) (Co-)Chain Maps:* Let $\mathcal{C} = (C_p, \partial_p)$ and $\widetilde{\mathcal{C}} = (\widetilde{C}_p, \widetilde{\partial}_p)$ be two chain complexes. A *chain map* $\phi : \mathcal{C} \to \widetilde{\mathcal{C}}$ is a $p$-family of homomorphisms

$$\phi_p : C_p \longrightarrow \widetilde{C}_p$$

such that $\widetilde{\partial}_p \circ \phi_p = \phi_{p-1} \circ \partial_p$, *i.e.*, the following diagram is *commutative*:

$$\begin{array}{ccc} C_p & \xrightarrow{\phi_p} & \widetilde{C}_p \\ \partial_p \downarrow & & \downarrow \widetilde{\partial}_p \\ C_{p-1} & \xrightarrow{\phi_{p-1}} & \widetilde{C}_{p-1} \end{array}$$

Cochain maps are defined analogously.

## III. Matrix Representation

In this section we introduce a block-matrix representation of the dual complexes $\mathcal{C} = (C_p, \partial_p)$ and $\mathcal{C}^* = (C^p, \delta^p)$ associated with a decomposition of the computational domain, and call it the *Hasse matrix*.

### A. Block-Matrix Decomposition

Let $K$ be a $d$-complex and $\mathcal{H}(K)$ its Hasse graph. The *Hasse matrix* $\mathbf{H}(K)$ is defined by the block structure shown in Fig. 5. Its transpose is the Hasse matrix of the dual complex $K^*$, whose Hasse graph $\mathcal{H}(K^*)$ is isomorphic to $\mathcal{H}(K)$, with $K_p^* \cong K_{d-p}$ ($0 \leq p \leq d$), where the boundary and coboundary operators are swapped by duality: $\mathbf{H}(K^*) = \mathbf{H}(K)^t$.

Since matrices representing boundary and coboundary operators are the transpose of each other, each block in the Hasse matrix may be interpreted either way. As shown in Fig. 5, we prefer to label blocks with *co*boundary matrices (and their transposes). The reason is that $\delta$'s acting on cochains are the discrete counterpart of the *exterior derivative* $d$ acting on *differential forms*. For typographic convenience, we shall occasionally write $\delta_p$ in lieu of $\delta^p$.

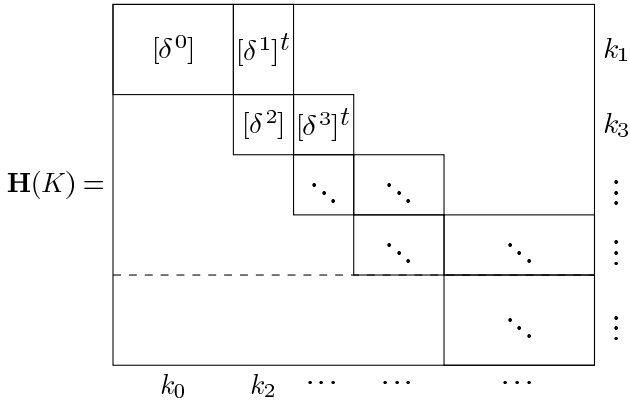

Fig. 5. Block structure of the Hasse matrix: the whole scheme holds for $d$ odd; for $d$ even, the last block-row should be discarded.

*Example 4 (Hasse Matrix of a 2-Complex):* Let $K$ be the 2-complex in Fig. 3, whose Hasse diagram is represented in Fig. 4. Its Hasse matrix is the $k_1 \times (k_0 + k_2) = 7 \times 8$ matrix

$$\mathbf{H}(K) = \begin{pmatrix} -.25 & .25 & 0 & 0 & 0 & 0 & 1. & 0 \\ -1. & 0 & 1. & 0 & 0 & 0 & -.25 & 0 \\ 0 & -1. & 0 & 1. & 0 & 0 & .25 & 0 \\ 0 & 0 & -.25 & .25 & 0 & 0 & -1. & .5 \\ 0 & 0 & -.5 & 0 & .5 & 0 & 0 & -.25 \\ 0 & 0 & 0 & -.5 & 0 & .5 & 0 & .25 \\ 0 & 0 & 0 & 0 & -.25 & .25 & 0 & -.5 \end{pmatrix}.$$

The left $k_1 \times k_0 = 7 \times 6$ block equals $[\delta^0] = [\partial_1]^t$; the right $k1 \times k_2 = 7 \times 2$ block equals $[\delta^1]^t = [\partial_2]$.

## IV. Euler Operators

In solid modeling it is common to refer to Euler operators as an independent set of operators that transform a boundary representation of a solid into a different one, satisfying the Euler-Poincaré formula [3], [15]. They may be allowed to change the Euler characteristic, whose definition is recalled below.

### A. Euler Characteristic

A well-known invariant associated with a finite $d$-dimensional cell complex $K$ is its *Euler characteristic* $\chi(K)$, defined as the alternating sum

$$\chi(K) = k_0 - k_1 + k_2 - k_3 + \cdots + (-1)^d k_d \,.$$

For polyhedra homeomorphic to the 2-sphere, the Euler characteristic is $k_0 - k_1 + k_2 = V - E + F = 2$.

According to the above, the *simplest* set of independent refining (coarsening) operators for a $d$-space that do not change its Euler characteristic has to increase (decrease) by one both $k_{p-1}$ and $k_p$, for $1 \leq p \leq d$. Therefore, there are $d$ elementary refining operators, and the same number of elementary coarsening operators.

*1) Properties of the Euler Characteristic:* The Euler characteristic is additive under *disjoint* union: if $M$ and $N$ are disjoint topological spaces ($M \cap N = \emptyset$), then

$$\chi(M \sqcup N) = \chi(M) + \chi(N).$$

More generally, if $M$ and $N$ are subspaces of a larger space $X$, then so are their union and intersection, and the Euler characteristic obeys the rule:

$$\chi(M \cup N) = \chi(M) + \chi(N) - \chi(M \cap N).$$

Moreover, the Euler characteristic of the product $M \times N$ equals the product of the Euler characteristics of $M$ and $N$:

$$\chi(M \times N) = \chi(M) \, \chi(N).$$

### B. Make and Kill Operators

The simplest Euler operators that transform a cell complex $K$ into another complex $\widetilde{K}$ such that $\chi(\widetilde{K}) = \chi(K)$ add (remove) one $p$-cell and one $(p+1)$-cell to (from) the complex. They will be denoted respectively by $\beta$ and $\kappa$, from the Greek words '*blastos*' (construction) and '*klastos*' (destruction).

By definition, the operator $\beta^p$ adds a $p$-cell and a $(p+1)$-cell to $K$, thus transforming it into $\widetilde{K}$. The reverse operator $\kappa^p$ deletes a $p$-cell and a $(p-1)$-cell. In the following, we provide detailed information on the construction of refining (or *make*) operators $\beta^p$ ($p = 0, \ldots, d-1$). Coarsening (or *kill*) operators $\kappa^p$ ($p = 1, \ldots, d$), which we will not elaborate on, are constructed analogously.

Let us first discuss how the whole hierarchy of coboundary operators transforms under the action of a make operator:

$$\beta^p : \delta_q \mapsto \widetilde{\delta}_q \qquad (q = 0, \ldots, n-1).$$

It is easily seen that $\beta^p$ affects in a nontrivial way only the coboundary operators whose domain and/or codomain change under its action, namely:

1) $\delta_{p+1} \mapsto \widetilde{\delta}_{p+1}$ ,
2) $\delta_{p-1} \mapsto \widetilde{\delta}_{p-1}$ ,
3) $\delta_p \mapsto \widetilde{\delta}_p$ ,

as shown by the following commutative diagram:

$$\widetilde{C}^{p+2} = C^{p+2} \xleftarrow{\delta_{p+1}} C^{p+1} \xleftarrow{\delta_p} C^p \xleftarrow{\delta_{p-1}} C^{p-1} = \widetilde{C}^{p-1}$$

as shown by the following commutative diagram (with $\widetilde{\delta}_{p+1}$, $\beta^p$, $\beta^p$, $\widetilde{\delta}_{p-1}$ arrows and $\widetilde{C}^{p+1}$, $\widetilde{C}^p$ with $\widetilde{\delta}_p$).

Three different computations have to be performed, depending on whether the domain only changes (case 1), or the codomain only (case 2), or both change (case 3).

*1) Addition of a Column $(\delta_{p+1} \mapsto \widetilde{\delta}_{p+1})$:* Let the matrix $[\delta_{p+1}]$ be $m \times n$; then, the matrix $[\widetilde{\delta}_{p+1}]$ will be $m \times (n+1)$. The column to be added to $[\delta_{p+1}]$ pinpoints the $(p+2)$-cells in $\widetilde{K}$ incident on the new cell $\widetilde{\sigma}_{p+1}$. It is a linear combination of the columns of $[\delta_{p+1}]$:

$$[\widetilde{\delta}_{p+1}] = [\delta_{p+1}] \left( \mathbf{I}_{n \times n} \, \middle| \, \begin{matrix} c_1 \\ \vdots \\ c_n \end{matrix} \right) = [\delta_{p+1}] \, \mathbf{C} \, .$$

*2) Addition of a Row $(\delta_{p-1} \mapsto \widetilde{\delta}_{p-1})$:* The row to be added to $[\delta_{p-1}]$ pinpoints the $(p-1)$-cells in $\widetilde{K}$ incident on the new cell $\widetilde{\sigma}_p$. It is a linear combination of the rows of $[\delta_{p-1}]$:

$$[\widetilde{\delta}_{p-1}] = \left( \frac{\mathbf{I}_{m \times m}}{r_1 \, \cdots \, r_m} \right) [\delta_{p-1}] = \mathbf{R} \, [\delta_{p-1}] \, .$$

*3) Addition of a Column and a Row $(\delta_p \mapsto \widetilde{\delta}_p)$:* One row of $[\delta_p]$ (one chain in $C_p$) is substituted by *two* rows (two chains in $\widetilde{C}_p$), whose components on the new cell $\widetilde{\sigma}_p$ sum up to zero. The matrix $[\widetilde{\delta}_p]$ is obtained as the sum

$$[\widetilde{\delta}_p] = \sum_{i=1}^{3} \mathbf{S}_i \, [\delta_p] \, \mathbf{T}_i \, ,$$

where the first term $(i=1)$ provides the contribution of the split cell $\sigma_{p+1}$, the second $(i=2)$ the contribution of the new cell $\widetilde{\sigma}_{p+1}$, and the third $(i=3)$ the contribution of all other $(p+1)$-cells in $K$.

### C. Examples

A minimal 2-complex $K$ is represented in Fig. 6. Cell sizes are assigned as follows: $\mu_1^1 = \sqrt{2}$, $\mu_1^2 = \mu_1^3 = 1.$, $\mu_2^1 = .5$. Fig. 8 shows its refinement $\widetilde{K}$, obtained by applying first the operator $\beta^0$ to halve the 1-cell $\sigma_1^1$, then the operator $\beta^1$ to halve the 2-cell $\sigma_2^1$. Fig. 7 shows the intermediate stage $(\widehat{K})$.

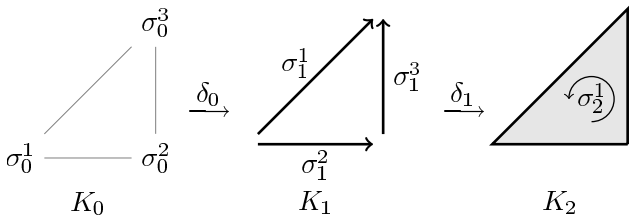

Fig. 6. Coarse complex $K \cong (K_0, K_1, K_2)$.

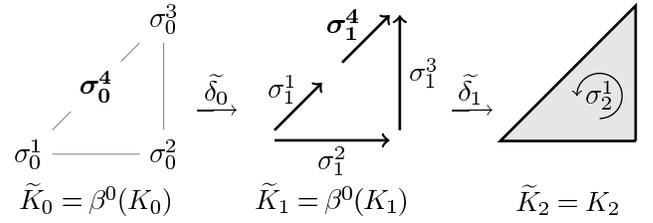

Fig. 7. First refining step: $\widehat{K} = \beta^0(K) \cong (K_0 \cup \{\sigma_0^4\}, K_1 \cup \{\sigma_1^4\}, K_2)$.

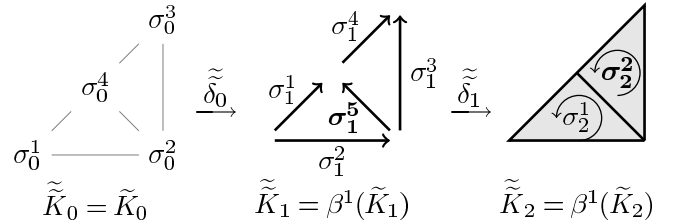

Fig. 8. Second refining step: $\widetilde{K} \cong \beta^1(\widehat{K}) = (\widehat{K}_0, \widehat{K}_1 \cup \{\sigma_1^5\}, \widehat{K}_2 \cup \{\sigma_2^2\})$.

*Example 5 (Coboundary $\delta_0 : C^0(K) \to C^1(K)$):* Domain and codomain have both dimension 3. From Fig. 6 it is seen that the matrix representation of $\delta_0$ is

$$[\delta_0] = \begin{pmatrix} -1/\sqrt{2} & 0 & 1/\sqrt{2} \\ -1. & 1. & 0 \\ 0 & -1. & 1. \end{pmatrix} \, .$$

*Example 6 (Coboundary $\delta_1 : C^1(K) \to C^2(K)$):* In this case, $k_1 = 3$ and $k_2 = 1$, so that

$$[\delta_1] = \begin{pmatrix} -2\sqrt{2} & 2. & 2. \end{pmatrix} \, .$$

*Example 7 (Coboundary $\widetilde{\delta}_0 : C^0(\widehat{K}) \to C^1(\widehat{K})$):* We have now $k_0 = k_1 = 3+1$. In Figs. 7, 8 the new cells are highlighted in boldface. Since both domain and codomain of $\widehat{\delta}_0$ are larger than those of $\delta_0$, the new coboundary operator is given by the sum of three contributions (see Section IV-B.3):

$$[\widetilde{\delta}_0] = \begin{pmatrix} \mathbf{S}_1 & \mathbf{S}_2 & \mathbf{S}_3 \end{pmatrix} [\delta_0] \begin{pmatrix} \mathbf{T}_1 \\ \mathbf{T}_2 \\ \mathbf{T}_3 \end{pmatrix} \, ,$$

where

$$\mathbf{S}_1 = \begin{pmatrix} 1 & 0 & 0 \\ 0 & 0 & 0 \\ 0 & 0 & 0 \\ 0 & 0 & 0 \end{pmatrix} \, , \quad \mathbf{S}_2 = \begin{pmatrix} 0 & 0 & 0 \\ 0 & 0 & 0 \\ 0 & 0 & 0 \\ 1 & 0 & 0 \end{pmatrix} \, ,$$

$$\mathbf{S}_3 = \begin{pmatrix} 0 & 0 & 0 \\ 0 & 1 & 0 \\ 0 & 0 & 1 \\ 0 & 0 & 0 \end{pmatrix} \, .$$

Matrices $\mathbf{S}_1$, $\mathbf{S}_2$ extract the row of $[\delta_0]$ that corresponds to the 1-cell $\sigma_1^1$ to be split (recall that a row of $[\delta_0]$ equals a column of $[\partial_1]$): $\mathbf{S}_1$ associates that row to $\widetilde{\sigma}_1^1$, while $\mathbf{S}_2$ associates it to the added cell $\widetilde{\sigma}_1^4$. Matrix $\mathbf{S}_3$ selects the remaining rows of $[\delta_0]$ and keeps them unchanged. Matrices $\mathbf{S}_i$ act on $[\delta_0]$ as

follows:

$$\mathbf{S}_1\left[\delta_0\right] = \begin{pmatrix} -\dfrac{1}{\sqrt{2}} & 0 & \dfrac{1}{\sqrt{2}} \\ 0 & 0 & 0 \\ 0 & 0 & 0 \\ 0 & 0 & 0 \end{pmatrix}, \quad \mathbf{S}_2\left[\delta_0\right] = \begin{pmatrix} 0 & 0 & 0 \\ 0 & 0 & 0 \\ 0 & 0 & 0 \\ -\dfrac{1}{\sqrt{2}} & 0 & \dfrac{1}{\sqrt{2}} \end{pmatrix},$$

$$\mathbf{S}_3\left[\delta_0\right] = \begin{pmatrix} 0 & 0 & 0 \\ -1. & 1. & 0 \\ 0 & -1. & 1. \\ 0 & 0 & 0 \end{pmatrix}.$$

Each column of a $\mathbf{T}_i$ matrix corresponds to a 1-cell in $\widetilde{K}$. Post-multiplication yields:

$$\mathbf{S}_1\left[\delta_0\right]\mathbf{T}_1 = \begin{pmatrix} -\sqrt{2} & 0 & 0 & \sqrt{2} \\ 0 & 0 & 0 & 0 \\ 0 & 0 & 0 & 0 \\ 0 & 0 & 0 & 0 \end{pmatrix},$$

$$\mathbf{S}_2\left[\delta_0\right]\mathbf{T}_2 = \begin{pmatrix} 0 & 0 & 0 & 0 \\ 0 & 0 & 0 & 0 \\ 0 & 0 & 0 & 0 \\ 0 & 0 & \sqrt{2} & -\sqrt{2} \end{pmatrix},$$

$$\mathbf{S}_3\left[\delta_0\right]\mathbf{T}_3 = \begin{pmatrix} 0 & 0 & 0 & 0 \\ -1. & 1. & 0 & 0 \\ 0 & -1. & 1. & 0 \\ 0 & 0 & 0 & 0 \end{pmatrix}.$$

Summing up, we obtain (see Fig. 7):

$$[\widetilde{\delta_0}] = \begin{pmatrix} -\sqrt{2} & 0 & 0 & \sqrt{2} \\ -1. & 1. & 0 & 0 \\ 0 & -1. & 1. & 0 \\ 0 & 0 & \sqrt{2} & -\sqrt{2} \end{pmatrix}.$$

*Example 8 (Coboundary $\widetilde{\delta_1} : C^1(\widetilde{K}) \to C^2(\widetilde{K})$):* In this case, $\widetilde{k}_1 = 3 + 1$ and $\widetilde{k}_2 = 1$. Hence, $[\widetilde{\delta_1}] = [\delta_1]\,\mathbf{C} =$

$$= [\delta_1] \begin{pmatrix} .5 & 0 & 0 & .5 \\ 0 & 1 & 0 & 0 \\ 0 & 0 & 1 & 0 \end{pmatrix} = \begin{pmatrix} -\sqrt{2} & 2. & 2. & -\sqrt{2} \end{pmatrix}.$$

*Example 9 (Coboundary $\widetilde{\widetilde{\delta_0}} : C^0(\widetilde{\widetilde{K}}) \to C^1(\widetilde{\widetilde{K}})$):* Here $\widetilde{\widetilde{k}}_0 = \widetilde{k}_0 = 4$, $\widetilde{\widetilde{k}}_1 = \widetilde{k}_1 + 1 = 5$, and we have: $[\widetilde{\widetilde{\delta_0}}] = \mathbf{R}\,[\widetilde{\delta_0}] =$

$$= \begin{pmatrix} 1 & 0 & 0 & 0 \\ 0 & 1 & 0 & 0 \\ 0 & 0 & 1 & 0 \\ 0 & 0 & 0 & 1 \\ 0 & 0 & \sqrt{2} & -1. \end{pmatrix} \begin{pmatrix} -\sqrt{2} & 0 & 0 & \sqrt{2} \\ -1. & 1. & 0 & 0 \\ 0 & -1. & 1. & 0 \\ 0 & 0 & \sqrt{2} & -\sqrt{2} \end{pmatrix}$$

$$= \begin{pmatrix} -\sqrt{2} & 0 & 0 & \sqrt{2} \\ -1. & 1. & 0 & 0 \\ 0 & -1. & 1. & 0 \\ 0 & 0 & \sqrt{2} & -\sqrt{2} \\ 0 & -\sqrt{2} & 0 & \sqrt{2} \end{pmatrix}.$$

*Example 10 (Coboundary $\widetilde{\widetilde{\delta_1}} : C^1(\widetilde{\widetilde{K}}) \to C^2(\widetilde{\widetilde{K}})$):* Here $\widetilde{\widetilde{k}}_1 = \widetilde{k}_1 + 1 = 5$ and $\widetilde{\widetilde{k}}_2 = \widetilde{k}_2 + 1 = 2$. By performing the

same operations as in Example 7, we get (see Fig. 8):

$$[\widetilde{\widetilde{\delta_1}}] = \begin{pmatrix} \mathbf{S}_1 & \mathbf{S}_2 & \mathbf{S}_3 \end{pmatrix}[\widetilde{\delta_1}] \begin{pmatrix} \mathbf{T}_1 \\ \mathbf{T}_2 \\ \mathbf{T}_3 \end{pmatrix}$$

$$= \begin{pmatrix} -2\sqrt{2} & 4. & 0 & 0 & 2\sqrt{2} \\ 0 & 0 & 4. & -2\sqrt{2} & -2\sqrt{2} \end{pmatrix}.$$

## V. Hasse Transformations

Let $K$ be a $d$-complex and $\mathbf{H}(K)$ its $n \times m$ Hasse matrix. Notice that $\chi(K) = m - n$ so that, while both $m$ and $n$ increase under topology-preserving refinements, their difference does not. We introduce the *Hasse transformation*

$$\beta^p : \mathcal{M}_m^n \to \mathcal{M}_{m+1}^{n+1}$$
$$\mathbf{H}(K) \mapsto \mathbf{H}(\widetilde{K})$$

induced by the make operator $\beta^p$ that splits the $(p+1)$-cell $\sigma_{p+1}^h$ into two cells, namely

$$\widetilde{\sigma}_{p+1}^h \quad \text{and} \quad \widetilde{\sigma}_{p+1}^{(k_{p+1})+1},$$

and adds a new $p$-cell, namely

$$\widetilde{\sigma}_p^{k_p+1}.$$

Let us assume—to be specific, but without loss of generality—that $d = 3$. In this case, $\mathbf{H}(K)$ is comprised of two diagonal blocks, $[\delta_0]$ and $[\delta_2]$, and one upper-diagonal block, $[\delta_1]^t$ (see Example 4 in Section III).

*Example 11 (Make Operators in 3D):* Three different algorithmic patterns arise, depending on the order of $\beta^p$:

$$\beta^0(\mathbf{H}(K)) = \left( \begin{array}{c|c} \begin{pmatrix} \mathbf{S}_1 & \mathbf{S}_2 & \mathbf{S}_3 \end{pmatrix}[\delta_0]\begin{pmatrix} \mathbf{T}_1 \\ \mathbf{T}_2 \\ \mathbf{T}_3 \end{pmatrix} & \mathbf{R}\,[\delta_1]^t \\ \hline \mathbf{0} & [\delta_2] \end{array} \right),$$

$$\beta^1(\mathbf{H}(K)) = \left( \begin{array}{c|c} \mathbf{R}[\delta_0] & \begin{pmatrix} \mathbf{S}_1 & \mathbf{S}_2 & \mathbf{S}_3 \end{pmatrix}[\delta_1]^t \begin{pmatrix} \mathbf{T}_1 \\ \mathbf{T}_2 \\ \mathbf{T}_3 \end{pmatrix} \\ \hline \mathbf{0} & [\delta_2]\mathbf{C} \end{array} \right),$$

$$\beta^2(\mathbf{H}(K)) = \left( \begin{array}{c|c} [\delta_0] & [\delta_1]^t\mathbf{C} \\ \hline \mathbf{0} & \begin{pmatrix} \mathbf{S}_1 & \mathbf{S}_2 & \mathbf{S}_3 \end{pmatrix}[\delta_2]\begin{pmatrix} \mathbf{T}_1 \\ \mathbf{T}_2 \\ \mathbf{T}_3 \end{pmatrix} \end{array} \right).$$

## VI. Hyperplane Splitting

In this section we discuss the SPLIT subdivision algorithm introduced in [29], rephrasing it in terms of the algebraic machinery developed in the previous sections. Subdivision algorithms allow to represent a smooth surface via the generation of coarser piecewise-linear meshes. They are the result of an iterative process of subdividing each polygonal face into smaller faces that better approximate the smooth surface. Subdivision is also used in combination with multigrid algorithms, for generation of finer 3D domain meshes in computational science.

The algorithm [29] works efficiently on a single $d$-cell of a $d$-complex, by locally splitting it with a hyperplane. The SPLIT algorithm performs first a numerical computation classifying all the vertices with respect to a given hyperplane, then carries out symbolic manipulations on higher dimensional cells, in order to obtain the resulting topology. Our algebraic formulation is general and easy to implement using standard packages for sparse-matrix computation [30]. The SPLIT algorithm is a useful tool for refining cell complexes, providing the ability to compute Boolean operations when combined with BSP trees in a progressive way [27]. The SPLIT algorithm is also useful to approximate continuous maps between cell complexes.

Subdivisions of cell complexes are formally defined as follows [13]: $\widetilde{K}$ is a *subdivision* of $K$ if
1) for each $\widetilde{\sigma} \in \widetilde{K}$, there exists $\sigma \in K$ such that $\widetilde{\sigma} \subseteq \sigma$;
2) for each $\sigma \in K$, there exists a finite subset $\{\widetilde{\sigma}_i\} \subseteq \widetilde{K}$, such that $\sigma = \cup_i \widetilde{\sigma}_i$.

The SPLIT algorithm—as detailed in the following Section VI-A—satisfies Property 1 by construction. Property 2 is also satisfied, since every cell in $K$ is mapped into the union of at most two cells ($\widetilde{\sigma}^-$ and $\widetilde{\sigma}^+$) in $\widetilde{K}$.

### A. The SPLIT Algorithm

Let us first introduce two auxiliary operators, useful for the matrix formulation of the SPLIT algorithm.

*Definition 1 (Sign):* The function $\mathrm{sgn}_\varepsilon : \mathbb{R}^d \to \{-1, 0, 1\}^d$ returns the matrix listing the signs of the elements $v_i$ of a $d$-tuple $\mathbf{v} = (v_i)$, accounting for a numerical tolerance $\varepsilon > 0$:

$$(\mathrm{sgn}_\varepsilon \mathbf{v})_i = \begin{cases} -1, & v_i < -\varepsilon \\ 0, & -\varepsilon \leq v_i \leq \varepsilon \\ 1, & v_i > \varepsilon \end{cases}$$

*Definition 2 (Absolute value):* The function abs acts on a real-valued matrix $\mathbf{M} = (m_{ij})$, returning the matrix of the absolute values of its elements:

$$\mathrm{abs}\,(m_{ij}) = (|m_{ij}|).$$

Let the splitting hyperplane

$$\mathfrak{h} = \left\{ (x_1, \ldots, x_d) \in \mathbb{E}^d \,\middle|\, \sum_i h^i x_i = b \right\}$$

be identified with the linear (*i.e.*, affine homogeneous) form $\mathbb{E}^{d+1} \to \mathbb{R}$ characterized by the row matrix

$$\mathbf{h} = \begin{pmatrix} h^1 & h^2 & \ldots & h^d & -b \end{pmatrix}.$$

Let $\mathbf{v}$ be the column matrix listing the homogeneous coordinates of the 0-cell $\sigma_0$:

$$\mathbf{x} = \begin{pmatrix} x_1 & x_2 & \ldots & x_d & 1 \end{pmatrix}^t.$$

Clearly, $\sigma_0$ belongs to the *above* (*below*) subspace $\mathfrak{h}^+$ ($\mathfrak{h}^-$) if and only if $\mathbf{h}\,\mathbf{x} < 0$: computing the sign of the scalar product $\mathbf{h}\,\mathbf{x}$ solves the *point location problem*. After introducing the matrix

$$\mathbf{X} := \begin{pmatrix} \mathbf{x}_1 & \mathbf{x}_2 & \cdots & \mathbf{x}_{k_0} \end{pmatrix},$$

collecting the homogeneous coordinates of all the 0-cells in $K_0$, these are classified with respect to the $\mathfrak{h}$ splitting hyperplane by the *classification function* $c^0 : K_0 \to \{-1, 0, 1\}$ represented by the row matrix

$$\mathbf{c}^0 := \mathrm{sgn}_\varepsilon(\mathbf{h}\,\mathbf{X}).$$

The SPLIT algorithm proceeds hierarchically from 0-cells up to $d$-cells by (a) classifying the cells with respect to the splitting hyperplane, and (b) updating the cell complex accordingly, including the new elements in the skeletons of all orders. The algorithm is sketched in Fig. 9. The inequality in step 5 characterizes the $p$-cells that intersect the splitting hyperplane $\mathfrak{h}$.

---

**algorithm** SPLIT (**input**: $K, \mathbf{X}, \mathbf{h}$; **output**: $\widetilde{K}, \widetilde{\mathbf{X}}$);

1) $p := 0$
2) Classify 0-cells: $\mathbf{c}^0 := \mathrm{sgn}_\varepsilon(\mathbf{h}\,\mathbf{X})$
3) $p := p + 1$
4) Classify $p$-cells:
   $\mathbf{c}^p := (\mathrm{abs}\,\mathbf{B}_{p-1}^t)\,\mathbf{c}^{p-1}$
   $\mathbf{a}^p := (\mathrm{abs}\,\mathbf{B}_{p-1}^t)\,\mathrm{abs}\,\mathbf{c}^{p-1}$
5) **foreach** $|c_i^p| \neq a_i^p$ **do**: Update the cell complex:
   split the $i$-th $p$-cell: $K := \beta^{p-1}(K)$;
   set the new element value: $c_{k_{p-1}}^{p-1} := 0$
6) Re-classify the $p$-cells of the updated cell complex:
   $\mathbf{c}^p := \mathrm{sgn}_\varepsilon\big((\mathrm{abs}\,\mathbf{B}_{p-1}^t)\,\mathbf{c}^{p-1}\big)$
7) **if** $p < d$ **then** GOTO step 3, **else** STOP.

---

Fig. 9. The SPLIT algorithm, implemented by using classification functions and the $p$-incidence matrices introduced in Section II-B.4.

### B. Splitting Example

Let us go back to the 2-complex considered in Section IV-C, and refine it with the splitting hyperplane specified in Fig. 10(a). The reader should recall Figs. 6–8 and refer to them to locate cells by name.

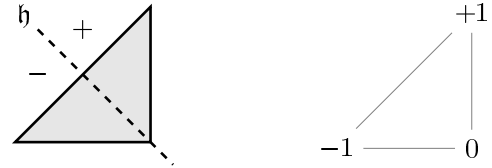

Fig. 10. (a) The splitting hyperplane $\mathfrak{h}$; (b) the classification of 0-cells.

The SPLIT algorithm is initialized by setting $p = 0$ and classifying 0-cells:

$$\mathbf{c}^0 = \mathrm{sgn}_\varepsilon\big(\mathbf{h}\begin{pmatrix} \mathbf{x}_1 & \mathbf{x}_2 & \mathbf{x}_3 \end{pmatrix}\big) = \begin{pmatrix} -1 & 0 & 1 \end{pmatrix},$$

as shown in Fig. 10(b). Then, $p$ is increased to 1 and 1-cells are classified by computing:

$$\mathbf{c}^1 = (\mathrm{abs}\,\mathbf{B}_0^t)\,\mathbf{c}^0 = \begin{pmatrix} 0 & -1 & 1 \end{pmatrix},$$

$$\mathbf{a}^1 = (\mathrm{abs}\,\mathbf{B}_0^t)\,\mathrm{abs}\,\mathbf{c}^0 = \begin{pmatrix} 2 & 1 & 1 \end{pmatrix}.$$

The result is illustrated in Fig. 11: we see that $\sigma_1^1$ should be split, since $|c_1^1| \neq a_1^1$. The $\beta^0$ operator adds a new 0-cell (classified to 0) and a new 1-cell (see Fig. 12). The two 1-cells resulting from the split are reclassified, as shown in Fig. 12. Then, $p$ is increased to 2 and 2-cells are classified:

$$\mathbf{c}^2 = (\text{abs } \mathbf{B}_1^t)\, \mathbf{c}^1 = \begin{pmatrix} 1 & 1 & 1 & 1 \end{pmatrix} \begin{pmatrix} -1 & -1 & 1 & 1 \end{pmatrix}^t = 0\,,$$

$$\mathbf{a}^2 = (\text{abs } \mathbf{B}_1^t)\, \text{abs } \mathbf{c}^1 = \begin{pmatrix} 1 & 1 & 1 & 1 \end{pmatrix} \begin{pmatrix} 1 & 1 & 1 & 1 \end{pmatrix}^t = 4\,,$$

(see Fig. 13a). Hence, $\sigma_2^1$ gets split, the splitting being executed by the $\beta^1$ operator that creates one 1-cell and one 2-cell as shown in Fig. 13. Finally, 2-cells are reclassified and the algorithm terminated, since $p = d$. The result is illustrated in Fig. 13(c).

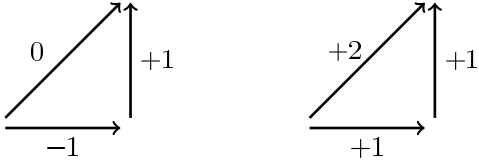

Fig. 11. The classification functions $\mathbf{c}^1$ and $\mathbf{a}^1$ used to detect 1-cells intersecting the splitting hyperplane.

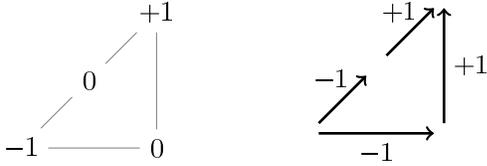

Fig. 12. The updated cell complex, with 1-cells reclassified.

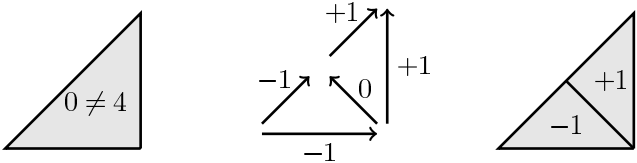

Fig. 13. (a) Classification of 2-cells; (b) the classification function on the refined 1-skeleton; and (c) the refined 2-skeleton with the classification of 2-cells.

### C. Subdivision of a Complex

Since SPLIT is a subdivision generator, the process can be iterated any number of times: for any $n \in \mathbb{N}$, $(\text{SPLIT}_n \circ \cdots \circ \text{SPLIT}_1)K$ is a subdivision of the complex $K$, which we shall call an $n$-iterated subdivision and denote briefly by $\text{SPLIT}^n K$.

From the finite approximation theorem [13] the following result follows. Let $\varphi : [\![K]\!] \to [\![L]\!]$ be any continuous map from the support of a finite complex $K$ to the support of a complex $L$ (recall from Example 2 that $[\![K]\!]$ is the point set union of all cells in $K$). Then, for an $n$-iterated subdivision of $K$ with $n$ big enough, there exists a map $\psi : \text{SPLIT}^n K \to L$ approximating $\varphi$.

Another important property of the SPLIT subdivision is guaranteed by the algebraic subdivision theorem [13]. The splitting induces a unique chain map (as defined in Section II-C.2) $\zeta : K \to \widetilde{K} := \text{SPLIT } K$. Consequently, boundaries in the refined cell complex $\widetilde{K}$ may be evaluated by applying the chain map $\zeta$ to boundaries in the coarse cell complex $K$.

## VII. Adjacency and Laplace-deRham Operators

In graph theory, the adjacency matrix of vertices is one of the many representations of a graph, i.e., a 1-complex $K \cong (K_0, K_1)$. The well-known relation between incidence and adjacency matrices of a graph can be extended to incidence and adjacency matrices of all orders $p \leq d$ of a general $d$-complex, for any $d \in \mathbb{N}$.

Definition 3: Let $\mathbf{M}_p$ be the measured $p$-incidence matrix introduced in Section II-B.5. Post- or pre-multiply $\mathbf{M}_p$ by its transpose, obtaining respectively:

$$\mathbf{A}_p^+ := \mathbf{M}_p \mathbf{M}_p^t\,, \qquad \mathbf{A}_{p+1}^- := \mathbf{M}_p^t \mathbf{M}_p\,. \tag{7}$$

The (symmetric) matrix $\mathbf{A}_p^+$ is, by definition, the adjacency matrix between $p$-chains through $(p+1)$-chains; analogously, $\mathbf{A}_{p+1}^-$ is the adjacency matrix between $(p+1)$-chains through $p$-chains.

As established by (5) and (6), $\mathbf{M}_p$ represents the boundary operator $\partial_{p+1}$ with respect to the standard bases of $C_{p+1}$ and $C_p$, and its transpose $\mathbf{M}_p^t$ the coboundary operator $\delta^p$ with respect to the dual bases of $C^p$ and $C^{p+1}$. Therefore, their products in (7), while legitimate as matrix operations, cannot possibly represent products of boundary and coboundary operators, unless chains are identified with cochains. This identification is performed by introducing a sequence of linear isomorphisms $G_p$ ($0 \leq p \leq d$) between chain spaces and their dual cochain spaces:

$$
\begin{array}{ccccccccccccc}
\cdots & \xleftarrow{\delta^{p+1}} & C^{p+1} & \xleftarrow{\delta^p} & C^p & \xleftarrow{\delta^{p-1}} & C^{p-1} & \xleftarrow{\delta^{p-2}} & \cdots & \longleftarrow & C^1 & \xleftarrow{\delta^0} & C^0 \\
& & \big\uparrow{\scriptstyle G_{p+1}} & & \big\uparrow{\scriptstyle G_p} & & \big\uparrow{\scriptstyle G_{p-1}} & & & & \big\uparrow{\scriptstyle G_1} & & \big\uparrow{\scriptstyle G_0} \\
\cdots & \xrightarrow{\partial_{p+2}} & C_{p+1} & \xrightarrow{\partial_{p+1}} & C_p & \xrightarrow{\partial_p} & C_{p-1} & \xrightarrow{\partial_{p-1}} & \cdots & \longrightarrow & C_1 & \xrightarrow{\partial_1} & C_0\,.
\end{array}
$$

The (nondegenerate) bilinear form induced by $G_p$ on $C_p$ will be denoted by the same symbol: for all $c_p, c_p' \in C_p$,

$$G_p(c_p, c_p') := \langle G_p c_p, c_p' \rangle\,. \tag{8}$$

The inner product (8) is assumed to be symmetric: for all $c_p, c_p' \in C_p$, $G_p(c_p', c_p) = G_p(c_p, c_p') \Leftrightarrow G_p^* = G_p$.

The adjoint (or transpose) $\partial_{p+1}^\top$ of the boundary operator $\partial_{p+1}$ is characterized by the property:

$$G_p(c_p, \partial_{p+1} c_{p+1}) = G_{p+1}(\partial_{p+1}^\top c_p, c_{p+1})\,,$$

to be satisfied for all $c_p \in C_p$ and $c_{p+1} \in C_{p+1}$. It is easily checked that

$$\partial_{p+1}^\top = (G_{p+1})^{-1}\, \partial_{p+1}^*\, G_p = (G_{p+1})^{-1} \delta^p\, G_p\,.$$

The isomorphism $G_p$ is represented by the (symmetric) Gram matrix $\mathbf{G}_p$, whose entries $G_p^{ij}$ are the components of the $G_p$-images of the elements of the standard basis of $C_p$ in the dual basis of $C^p$:

$$G_p u_p^i = \sum_{j=1}^{k_p} G_p^{ij} \eta_j^p.$$

Consequently (recall (6)),

$$[\partial_{p+1}^\top] = \mathbf{G}_{p+1}^{-1} \mathbf{M}_p^t \mathbf{G}_p .$$

Therefore, the adjacency matrices introduced in (7) represent the (symmetric) *adjacency operators*

$$\partial_{p+1}\partial_{p+1}^\top , \quad \partial_{p+1}^\top \partial_{p+1}$$

only if the inner product hierarchy is *trivial*: for all $0 \le p \le d$,

$$\mathbf{G}_p = \mathbf{I}_{k_p \times k_p} \quad \Leftrightarrow \quad G_p u_p^i = \eta_i^p . \tag{9}$$

The discrete *Laplace-deRham operators*—a fundamental and ubiquitous ingredient of physical modeling—are defined as sums of *duals* of adjacency operators: for all $0 \le p \le d$,

$$\Delta_p := (\partial_{p+1}\partial_{p+1}^\top + \partial_p^\top \partial_p)^* = \delta_p^\top \delta_p + \delta_{p-1} \delta_{p-1}^\top ,$$

it being intended that $\partial_0$, $\delta_d$ (and hence $\delta_0^\top$, $\partial_d^\top$) are null. Of course, the straightforward representation

$$[\Delta_p] = \mathbf{A}_p^+ + \mathbf{A}_p^-$$

is only valid if the inner product hierarchy is trivial.

*Example 12 (Adjacency Matrices in 3D):* Let us consider the chain and cochain complexes associated with the 3-complex $K$ whose *topology* is given in Fig. 14. As far as *measure* and *inner product* structures are concerned, we make the trivial choices: $\mu_p^i = 1.$, $G_p u_p^i = \eta_i^p$ for all $0 \le p \le d$ and $1 \le i \le k_p$. As the reader will notice, these choices—though legitimate—do *not* mimic, not even roughly, the Euclidean properties of $[\![K]\!]$ suggested by Fig. 14.

Boundary and coboundary operators are represented by the matrices:

$$[\delta_0] = [\partial_1]^t = \begin{pmatrix} -1 & 0 & 0 & 1 & 0 \\ -1 & 0 & 1 & 0 & 0 \\ -1 & 1 & 0 & 0 & 0 \\ 0 & -1 & 0 & 1 & 0 \\ 0 & -1 & 1 & 0 & 0 \\ 0 & 0 & -1 & 1 & 0 \\ 0 & 0 & 0 & -1 & 1 \\ 0 & 0 & -1 & 0 & 1 \\ 0 & -1 & 0 & 0 & 1 \end{pmatrix},$$

$$[\delta_1] = [\partial_2]^t = \begin{pmatrix} -1 & 1 & 0 & 1 & 0 & 0 & 0 & 0 & 0 \\ 0 & -1 & 1 & 0 & 1 & 0 & 0 & 0 & 0 \\ 1 & 0 & -1 & 0 & 0 & -1 & 0 & 0 & 0 \\ 0 & 0 & 0 & -1 & -1 & 1 & 0 & 0 & 0 \\ 0 & 0 & 0 & -1 & 0 & 0 & -1 & 1 & 0 \\ 0 & 0 & 0 & 0 & -1 & 0 & -1 & 0 & 1 \\ 0 & 0 & 0 & 0 & 0 & 1 & 1 & 0 & -1 \end{pmatrix},$$

$$[\delta_2] = [\partial_3]^t = \begin{pmatrix} 1 & 1 & 1 & 1 & 0 & 0 & 0 \\ 0 & 0 & 0 & -1 & 1 & 1 & 1 \end{pmatrix}.$$

A straightforward computation yields the adjacency and Laplace-deRham matrices:

$$[\Delta_0] = [\partial_1]\,[\delta_0] = \begin{pmatrix} 3 & -1 & -1 & -1 & 0 \\ -1 & 4 & -1 & -1 & -1 \\ -1 & -1 & 4 & -1 & -1 \\ -1 & -1 & -1 & 4 & -1 \\ 0 & -1 & -1 & -1 & 3 \end{pmatrix},$$

$$[\delta_0]\,[\partial_1] = \begin{pmatrix} 2 & 1 & 1 & 1 & 0 & 1 & -1 & 0 & 0 \\ 1 & 2 & 1 & 0 & 1 & -1 & 0 & -1 & 0 \\ 1 & 1 & 2 & -1 & -1 & 0 & 0 & 0 & -1 \\ 1 & 0 & -1 & 2 & 1 & 1 & -1 & 0 & 1 \\ 0 & 1 & -1 & 1 & 2 & -1 & 0 & -1 & 1 \\ 1 & -1 & 0 & 1 & -1 & 2 & -1 & 1 & 0 \\ -1 & 0 & 0 & -1 & 0 & -1 & 2 & 1 & 1 \\ 0 & -1 & 0 & 0 & -1 & 1 & 1 & 2 & 1 \\ 0 & 0 & -1 & 1 & 1 & 0 & 1 & 1 & 2 \end{pmatrix},$$

$$[\partial_2]\,[\delta_1] = \begin{pmatrix} 2 & -1 & -1 & -1 & 0 & -1 & 0 & 0 & 0 \\ -1 & 2 & -1 & 1 & -1 & 0 & 0 & 0 & 0 \\ -1 & -1 & 2 & 0 & 1 & 1 & 0 & 0 & 0 \\ -1 & 1 & 0 & 3 & 1 & -1 & 1 & -1 & 0 \\ 0 & -1 & 1 & 1 & 3 & -1 & 0 & 1 & -1 \\ -1 & 0 & 1 & -1 & -1 & 3 & 1 & 0 & -1 \\ 0 & 0 & 0 & 1 & 0 & 1 & 2 & -1 & -1 \\ 0 & 0 & 0 & -1 & 1 & 0 & -1 & 2 & -1 \\ 0 & 0 & 0 & 0 & -1 & -1 & -1 & -1 & 2 \end{pmatrix},$$

$$[\partial_3]\,[\delta_2] = \begin{pmatrix} 1 & 1 & 1 & 1 & 0 & 0 & 0 \\ 1 & 1 & 1 & 1 & 0 & 0 & 0 \\ 1 & 1 & 1 & 1 & 0 & 0 & 0 \\ 1 & 1 & 1 & 2 & -1 & -1 & -1 \\ 0 & 0 & 0 & -1 & 1 & 1 & 1 \\ 0 & 0 & 0 & -1 & 1 & 1 & 1 \\ 0 & 0 & 0 & -1 & 1 & 1 & 1 \end{pmatrix},$$

$$[\delta_1]\,[\partial_2] = \begin{pmatrix} 3 & -1 & -1 & -1 & -1 & 0 & 0 \\ -1 & 3 & -1 & -1 & 0 & -1 & 0 \\ -1 & -1 & 3 & -1 & 0 & 0 & -1 \\ -1 & -1 & -1 & 3 & 1 & 1 & 1 \\ -1 & 0 & 0 & 1 & 3 & -1 & -1 \\ 0 & -1 & 0 & 1 & -1 & 3 & -1 \\ 0 & 0 & -1 & 1 & -1 & -1 & 3 \end{pmatrix},$$

$$[\Delta_3] = [\delta_2]\,[\partial_3] = \begin{pmatrix} 4 & -1 \\ -1 & 4 \end{pmatrix}.$$

The computation of $[\Delta_1]$ and $[\Delta_2]$ is left to the reader, who is invited to pay attention to the subtle cancellation of terms.

## VIII. GEOMETRY & PHYSICS MODELING

The real-valued (co)chain-complex formalism and the associated Hasse-matrix representation fit physical modeling in a natural and straightforward way. Chains may be tuned to endow cells with the physical *measures* of interest: length, area, volume, but also mass, charge, and so on. Cochains, on the other side, represent *densities* of all physical quantities associated with cells through *integration* with respect to those measures. The coboundary operator is behind the basic structural laws (balance and compatibility) involving physically meaningful cochains [17], [18], [31]. It is also well known that $p$-cochains are the discrete analogue of differential $p$-forms [32], [33]. Correspondingly, the cochain complex introduced in Section II-C.1 is the discrete precursor of the de Rham complex [34]–[36], naturally represented by the Hasse matrix (or its transpose). This view on physical modeling has been progressively advocated [22], [36], [37] as a way to increase numerical stability and accuracy of various numerical methods.

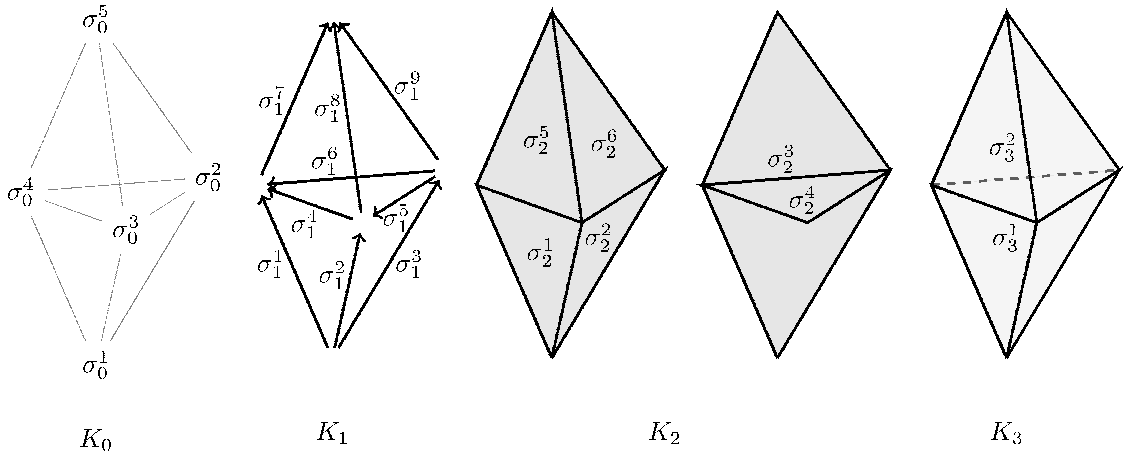

Fig. 14. A sample 3-complex $K \cong (K_0, K_1, K_2, K_3)$.

The Hasse matrix $\mathbf{H}(K)$, introduced in Section III-A, provides a compact representation of boundary $\partial$ and coboundary $\delta$ operators, acting respectively on chains or cochains defined on $K$. This representation depends only on: (i) the *topology* of the underlying cell complex, and (ii) the *measure* imparted to cells by chains.

A supplementary *inner-product* structure establishes a bridge between the chain and cochain complexes, as needed for introducing the adjacency and Laplace-deRham operators (cf. Section VII). The underlying topology stays untouched. Cell measures may—but need not—be related with the inner-product structure. The trivial choice (9) we made in Example 12 for the sake of argument, while expedient to use on any given $K$, is totally unrelated—in general—to the geometric properties of $[\![K]\!]$ relevant to the physics under consideration. Dealing properly with the identification between chains and cochains is essential for importing into the model the relevant, physics-based inner-product structure. This issue is also basic to gain the possibility of a meaningful information transfer from a cell complex $K$ to any of its refinements $\tilde{K}$ (and vice versa), and to establish a decent notion of convergence for refinement sequences.

A deeper discussion of this point—*i.e.*, how to construct physically meaningful inner products between chains—is beyond the scope of the present paper. Here, we just stress that the very same data structures and algorithms can be used concurrently for solid modeling and physics-based simulations.

From our vantage point, boundary representations and finite element meshes appear as two different aspects of the same Hasse representation. Furthermore, there is no fundamental distinction between different types of approximation methods, as we showed in [23], [24] for linear problems. Within our framework, the SPLIT algorithm described in Section VI-A becomes a powerful tool for progressive refinement not only of *shapes*, but also of the representation of *fields* living on those shapes.

## IX. Conclusion

Historically, the development of boundary representation schemes in solid modeling was driven by limited computational resources, and the usual space-time trade-offs [9]. Boundary representation was the solution of choice in order to (a) save memory, when RAM was small and expensive, and (b) spare disk access times, by answering topological queries efficiently.

Contrary to what might appear at first sight, the present approach does not imply higher theoretical complexity, since the number of non-zero elements in the Hasse matrix $\mathbf{H}(K)$ is essentially of the same order as the number of adjacency pointers in a typical graph-based representation of the cell complex $K$. Furthermore, the Hasse matrix serves as a unifying standard for all boundary representations, different graph structures corresponding to different methods [30] for encoding the nontrivial information contained in the sparse matrix $\mathbf{H}(K)$.

It should also be noted that chain complexes are a standard tool for representing and analyzing topological properties of arbitrary cellular spaces. It follows that the proposed Hasse matrix and Hasse transformations may codify much more general models, without restrictions on orientability, (co)dimension, manifoldness, connectivity, homology, and so on. The resulting framework, centered on a matrix representation of the domain of interest, unifies several geometric and physical finite formulations, and supports local progressive refinement and coarsening.

This approach is strongly motivated by the applications to be developed within the next generation of computational sciences, which demand large-scale simulation models of field problems where geometric and physical properties have to be generated, detailed, and refined *simultaneously* and *progressively*.

## Acknowledgment

The anonymous reviewers are heartily thanked for their remarks, which greatly helped us produce a better paper. Antonio DiCarlo acknowledges support from his own Department (DiS) and from MiUR (the Italian Ministry of University and Research) through the Project "Mathematical Models for Materials Science." Alberto Paoluzzi thanks IBM for the support provided through a Shared University Research grant. Research of Vadim Shapiro is supported in part by the US National Science Foundation grants CMMI-0500380, CMMI-0621116 and OCI-0636206.